\documentclass{slides}
\usepackage{amsmath}
\begin{document}
\title{
{\huge Cosmological Constant} \\or\\{\huge Cosmological
Potential\footnote{talk given at Conference "PROBLEMS OF VACUUM
ENERGY" Copenhagen, 23--26 August, 2000}}}
\author{
\vskip .2truecm
 {\large P.~Fiziev}
 \vskip .1truecm Theoretical Physics Department\\
 \vskip .1truecm Sofia University\\
 \vskip .1cm fiziev@phys.uni-sofia.bg
}
\maketitle
%

\newcommand{\lfrac}[2]{{#1}/{#2}}
\newcommand{\sfrac}[2]{{\small \hbox{${\frac {#1} {#2}}$}}}
\newcommand{\ben}{\begin{eqnarray}}
\newcommand{\een}{\end{eqnarray}}
\newcommand{\la}{\label}
%
Introduction: \vskip -.85truecm At present the original general
relativity (GR) is: \vskip -.85truecm A SUCCESSFUL THEORY of
gravity in description of gravitational phenomena at: \vskip
-.85truecm

$\bullet$ laboratory and earth surface scales,\\ $\bullet$ solar
system and star systems scales.\\ \vskip -.85truecm \vskip
-.85truecm QUITE GOOD in description of these phenomena: \vskip
-.85truecm

$\bullet$ at galaxies scales, and\\ $\bullet$ at the scales of the
whole Universe. \vskip -.85truecm PROBLEMATIC in description of:
\vskip -.85truecm

$\bullet$ rotation of galaxies and their motion in galactic
clausters\\ $\bullet$ initial singularity problem, early Universe
and inflation,\\ $\bullet$ recently discovered accelerated
expansion of the Universe,\\ $\bullet$ {\bf vacuum energy
problem}.

\newpage
The most promising modern theories like supergravity and
(super)string theories  incorporate naturally GR but at present:
\vskip -.5truecm

$\bullet$ are not developed enough to allow {\bf real}
experimental test,\\ $\bullet$ introduce large number of new
fields without {\bf real} physical basis.\vskip -.5truecm

Therefore it seems meaningful to look for some {\bf minimal
extension of GR} which: \vskip -.5truecm

$\bullet$ is compatible with known gravitational experiments,\\
$\bullet$ promises to overcome at least some of the problems,\\
$\bullet$ may be considered as a part of more general modern
theories.\vskip -.5truecm

Such {\bf minimal} extension must include one new scalar field
degree of freedom. Its contribution in the action of the theory
may be described in different (sometimes equivalent) ways, being
not fixed a priori.

\newpage
Here we outline the general properties of such model with one
additional scalar field $\Phi$ which differs from known
inflationary models, being conformal equivalent in scalar-tensor
sector to some quintessence models. We call

{\bf MINIMAL DILATONIC GRAVITY (MDG)}

the scalar-tensor model of gravity with action

\ben
{\cal A}_{G,\Lambda}=-{\sfrac c {2\bar\kappa}}\int d^4 x\sqrt{|g|}
\Phi \bigl( R + 2 \Lambda \Pi(\Phi) \bigr),
\la{A_Gc}
\een

$\bullet$ Branse-Dicke parameter $\omega(\Phi)\!\equiv\!0$ ({\bf
i.e., without standard kinetic term for $\Phi$ !}),

$\bullet$ {\bf cosmological constant} $\Lambda$ and

$\bullet$ {\bf dimensionless cosmological factor} $\Pi(\Phi)$.

The matter action ${\cal A}_M$ and matter equations of motion will
have usual GR form.

\newpage
Equations for metric $g_{\alpha\beta}$ and dilaton field $\Phi$:

\ben
\Phi \left(G_{\alpha\beta}\!-\!\Lambda \Pi(\Phi)
g_{\alpha\beta}\right) \!-\!(\nabla_\alpha
\nabla_\beta\!-\!g_{\alpha\beta}{\sqcap\hskip -.21in\sqcup})\Phi
\!=\! {\sfrac {\bar\kappa} {c^2}} T_{\alpha\beta}, \nonumber
\een
\ben{\sqcap\hskip -.21in\sqcup}\Phi\!+\!\Lambda {\sfrac
{dV}{d\Phi}}(\Phi)=\! {\sfrac {\bar\kappa} {3 c^2}}T.
\la{FEq}
\een

yield usual energy-momentum conservation law:
$$\nabla_\alpha\,T^\alpha_\beta=0.$$ In addition we have the
important relation:

\ben
R+2\Lambda \bigl(\Phi{\sfrac{d\Pi}{d\Phi}}(\Phi)
+\Pi(\Phi)\bigr)=0.
\la{RV}
\een

In (\ref{FEq}) the quantity: $${\sfrac {dV}{d\Phi}}={\sfrac 2 3}
\Phi\left(\Phi{\sfrac{d\Pi}{d\Phi}}-\Pi\right)$$ introduces {\bf
dilatonic potential} $V(\Phi)$.

It is convenient to introduce, too, {\bf a cosmological
potential}: $$U(\Phi)=\Phi \Pi(\Phi).$$
\newpage
Investigation of MDG was started by O'Hanlon (PRL,1972) in
connection with Fujii's theory of massive dilaton, but {\bf
without} any relation with cosmological constant problem.

In our special scalar-tensor model of gravity the cosmological
factor $\Pi(\Phi)$ (or the cosmological potential $U(\Phi)$) is
the {\bf only unknown function} which has to be chosen:

$\bullet$ to comply with {\bf all gravitational experiments and
observations}\\ $\bullet$ to solve the following {\bf inverse
cosmological problem}:

Determination of the factor $\Pi(\Phi)$ which yields given time
evolution of the scale parameter $A(t)$ in Robertson-Walker (RW)
model of Universe.

The action (\ref{A_Gc}) of MDG is a Helmholz action of nonlinear
gravity with lagrangian $$L_{NLG} \sim \sqrt{|g|} f(R),$$ or a
(4d) low energy limit of superstringy action (for metric and
dilaton only) in some {\bf new frame}. It follows from stringy
lagrangian $$L_{String} \sim \sqrt{|g|}e^{-2\phi}\left(R + 4
(\partial \phi)^2 +V_{SUSY}(\phi)\right)$$ after conformal
transformation (in $D$-dimensions): $$g_{\mu\nu} \rightarrow
e^{\pm {4\phi}/ \sqrt{(D-2)(D-1)} }g_{\mu\nu},$$ $$\phi
\rightarrow \Phi = \exp\left(-2\phi\left( 1 \pm \sqrt{(D-2)/
(D-1)}\,\right)\right).$$

The potential $V_{SUSY}(\phi)$ may originate from SUSY breaking
due to gaugino condensation, or may appear in more complicated
way. At present its form is not known exactly.

 The action (\ref{A_Gc}) appears, too, in a new model of gravity
with torsion and unusual local conformal symmetry after its
breaking in metric-dilaton sector only (PF: gr-qc/9809001).

\newpage

An {\bf essential new element} of our MDG (PF: gr-qc/9911037) is
the {\bf nonzero cosmological constant $\Lambda$.}

Nevertheless at present still exist doubts in astrophysical data:
$$\Omega_{\Lambda}\!=\!.65\pm .13,\,\,\,
\,\,\,\,\,\,H_0\!=\!(65\pm 5)\,km\,s^{-1}Mps^{-1}$$ \noindent
which determine $$\Lambda^{obs}=3\Omega_\Lambda H_0^2 c^{-2}= (.98
\pm .34) \times 10^{-56} cm^{-2}$$ we accept this observed value
of cosmological constant {\bf as a basic quantity} which  {\bf
defines natural units} for all other cosmological quantities,
namely:

$\bullet$ cosmological length: $$A_c = 1/\sqrt{\Lambda^{obs}} =
(1.02\pm .18)\times 10^{28} cm,$$

$\bullet$ cosmological time: $$T_c:=A_c/c= (3.4\pm .6)\times
10^{17} s=(10.8 \pm 1.9) Gyr,$$

$\bullet$ cosmological energy density: $$\varepsilon_c:={\frac
{\Lambda c^2} {\kappa}}= (1.16\pm .41)\times 10^{-7}
g~cm^{-1}~s^{-2},$$

$\bullet$ cosmological energy: $$E_c:=3 A_c^3\varepsilon_c=
3\Lambda^{-1/2} c^2 \kappa^{-1} =(3.7\pm .7) \times 10^{77} erg,$$

$\bullet$ cosmological momentum: $$P_c:= 3c / (\kappa
\sqrt{\Lambda^{obs}}) = (1.2 \pm .2)\times 10^{67} g~cm~s^{-1}$$

$\bullet$ and {\bf cosmological unit for action}: $${\cal A}_c:=
3c/(\kappa\Lambda^{obs})=(1.2 \pm .4)\times 10^{122}\,\hbar,$$

\noindent $\kappa$ being Einstein constant.

Further we use dimensionless variables like:

$\tau := t/T_c$, $a := A/A_c$, $h:= H\,T_c$ ($H:=A^{-1}dA/dt$
being Hubble parameter), $\epsilon_c = \varepsilon_c /
|\varepsilon_c|=\pm 1$, $\epsilon:=\varepsilon /
|\varepsilon_c|$-matter energy density, etc.

\newpage

{\Large Solar System and Earth-Surface Gravitational Experiments}

Properties of cosmological factor $\Pi(\Phi)$ derived from known
gravitational experiments:\vskip -.5truecm

1.The MDG with $\Lambda=0$ contradicts to solar system
gravitational experiments. The cosmological term $\Lambda\Pi(\Phi)
\neq 0$ in action (\ref{A_Gc}) is needed to overcome this
problem.\vskip -.5truecm

2. In contrast to O'Hanlon's model we wish MDG to reproduce GR
with $\Lambda\!\neq\!0$ for some
$$\Phi\!=\!\bar\Phi\!=\!const\neq\!0,$$ i.e., we consider a MDG
solution which coincides with original de Sitter solution.

Then from action (\ref{A_Gc}) we obtain {\bf normalization
condition} for cosmological factor $$\Pi(\bar\Phi)=1$$ and
Einstein constant $$\kappa=\bar\kappa/\bar\Phi.$$
\newpage
In vacuum, far from matter MDG have to allow {\bf week field
approximation} ($|\zeta|\ll 1$): $$\Phi=\bar\Phi(1+\zeta).$$ Then
the {\bf linearized} dilaton equation (\ref{FEq}): $${\sqcap\hskip
-.2in \sqcup} \zeta+ \zeta/l_\Phi^2\!=\!{\sfrac {\kappa}{3
c^2}}T$$ gives $${\frac{d\Pi}{d\Phi}}(\bar\Phi)=1/\bar\Phi.$$
Taylor series expansion of the function
${\sfrac{dV}{d\Phi}}(\Phi)$ around the value $\bar\Phi$ gives
relation $${\frac {d^2\Pi}{d\Phi^2}}(\bar\Phi)\!=\!{\sfrac 3
2}p^{-2}\bar\Phi^{-2}.$$ Then $$\Pi\!=\!1\!+ \zeta\!+{\frac 3 {4
p^2 }}\zeta^2\!+O(\zeta^3),$$ $$p\!=\!{\frac {l_\Phi}{A_c}}$$
\noindent being dimensionless Compton length of dilaton in
cosmological units.

4. {\bf Point particles of masses $m_a$} as source of metric and
dilaton fields give in Newtonian approximation gravitational
potential $\varphi({\bf r})$ and dilaton field $\Phi({\bf r})$:

\ben \varphi({\bf r})/c^2\!=\! - {\sfrac G {c^2}}\!\sum_a\!{\sfrac
{m_a}{|{\bf r - r}_a|}}\! \left(\!1\!+\!\alpha(p) e^{-|{\bf r -
r}_a|/l_\Phi} \right) \nonumber\\ - {\sfrac 1 6}p^2
\sum_a\!{\sfrac{m_a} M}\left(|{\bf r - r}_a|/l_\Phi\right)^{\!2},
\een
\ben \!\Phi({\bf r})/\bar\Phi\!=\!1+\!{\sfrac 2 3} {\sfrac
G{c^2(1-{\sfrac 4 3} p^2)}} \sum_a\!{\sfrac {m_a}{|{\bf r -
r}_a|}} e^{- |{\bf r - r}_a|/l_\Phi}, \la{SolNewton} \een
$G\!=\!{\sfrac {\kappa c^2} {8\pi}}(1\!-\!{\frac 4 3}p^2)$ is
Newton constant, $M\!=\!\sum_a m_a$. The term $$-{\sfrac 1 6}p^2\!
\sum_a{\sfrac{m_a} M}\left(|{\bf r\!-\!r}_a|/l_\Phi\right)^2\!=\!
-{\sfrac 1 6} \Lambda|{\bf r}-\!\sum_a \!{\sfrac {m_a} M}{\bf
r}_a|^2\!+\!const$$ in $\varphi$ is known from GR with $\Lambda
\neq 0$. It represents an {\bf universal anty-gravitational
interaction} of test mass $m$ with any mass $m_a$ via repellent
elastic force
\ben {\bf F}_{\!{}_\Lambda\,a}={\sfrac 1 3}\Lambda m c^2
{\sfrac{m_a} M}({\bf r - r}_a). \la{Fel} \een

\newpage
For {\bf solar system distances} $l\leq 1000 AU$ neglecting the
$\Lambda$ term ( of order $\leq 10^{-24}$) we compare the
gravitational potential $\varphi$ with specific MDG coefficient
$$\alpha(p)={\frac{1+4p^2}{3-4 p^2}}$$ with Cavendish type
experiments and obtain an experimental constraint $l_\Phi \leq
1.6$ [mm], or $$p < 2\times 10^{-29}.$$ Hence, in the solar system
the factor $e^{-l/l_\Phi }$ has a fantastic small values
($<\exp(-10^{13})$ for the Earth-Sun distances $l$, or $<
\exp(-3\times 10^{10})$ for the Earth-Moon distances $l$) and {\bf
there is no hope to find some differences between MDG and GR} in
this domain.

The corresponding constraint $$m_\Phi c^2 \geq 10^{-4}[eV]$$ does
not exclude a small value (with respect to the elementary
particles scales) for the rest energy of hypothetical
$\Phi$-particle.

\newpage
5. The {\bf parameterized-post-Newtonian(PPN) solution} of
equation (\ref{FEq}) is complicated, but because of the constraint
$p < 10^{-28}$ we may use with great precision Helbig's PPN
formalism (for $\alpha\!=\!{\sfrac 1 3}$). Because of the
condition $\omega\equiv 0$ we obtain much more definite
predictions then usual general relations between $\alpha$ and the
length $l_\Phi$:

$\bullet$ {\bf Nordtvedt Effect:}

In MDG a body with significant gravitational self-energy
$E_{{}_G}=\sum_{b\neq c} G{\sfrac {m_b m_c}{|{\bf r}_b - {\bf
r}_c|} }$ will not move along geodesics due to {\bf additional
universal anty-gravitational force}:
\ben
{\bf F}_{\!{}_N} = -{\sfrac 2 3} E_{{}_G}\nabla \Phi.
\la{NordtF}
\een
For usual bodies it is too small even at distances $|{\bf
r-r}_a|\!\leq\!l_\Phi$, because of the small factor $E_G$. Hence,
in MDG we have no strict strong equivalence principle nevertheless
{\bf the week equivalence principle is not violated}.

The experimental data for Nordtvedt effect,  caused by the Sun,
are formulated as a constraint $\eta=0 \pm .0015$  on the
parameter $\eta$ which in MDG becomes a function of the distance
$l$ to the source: $\eta(l)=-{\sfrac 1 2}\left(1+l/l_\Phi\right)
e^{-l/l_\Phi}$. This gives constraint $l_\Phi \leq 2\times 10^{10
}[m]$.

$\bullet$ {\bf Time Delay of Electromagnetic Waves}

The action of electromagnetic field does not depend on the field
$\Phi$. Therefore influence of $\Phi$ on the electromagnetic waves
in vacuum is possible only via influence of $\Phi$ on the
space-time metric. The solar system measurements of the time delay
of the electromagnetic pulses give the value $\gamma=1 \pm .001$
of this post Newtonian parameter. In MDG this yields the relation
$(1 \pm .001) g( l_{{}_{AU}})= 1$ and gives once more the
constraint $l_\Phi \leq 2\times 10^{10 }[m]$. Here
$g(l):=1+{\sfrac 1 3 }(1+l/l_\Phi) e^{-l/l_\Phi}$.

$\bullet$ {\bf Perihelion Shift}

For the perihelion shift of a planet orbiting around the Sun (with
mass $M_\odot$) in MDG we have: $$\delta \varphi = {\frac
{k(l_p)}{g(l_p)}}\delta\varphi_{{}_{GR}}.$$ Here $l_p$ is the
semimajor axis of the orbit of planet and $$k(l_p) \approx 1 +
{\sfrac 1 {18}}\left( 4 + {\sfrac{l_p^2}{l_\Phi^2}}{\sfrac {l_p
c^2}{GM_\odot}}\right) e^{-l_p/l_\Phi} -{\sfrac 1 {27}}
e^{-2l_p/l_\Phi}$$ is obtained neglecting its eccentricity. The
observed value of perihelion shift of Mercury gives the constraint
$l_\Phi \leq  10^{9 }[m]$.

{\em Conclusion:}

 {\bf In presence of dilaton field $\Phi$ are
impossible essential deviations from GR in solar system.

Observable deviations from Newton law of gravity may not be
expected at distances greater then few mm}.

\newpage

{\bf\large Vacuum Energy and True Vacuum Solution in MDG }

Total (true) tensor of energy momentum is:
\ben
TT_{\mu\nu}:= T_{\mu\nu}+ <\rho_0>c^2 g_{\mu\nu},
\la{TT}
\een
$<\rho_0>$ being the averaged energy density of zero quantum
fluctuations. For {\bf true vacuum solution} of MDG:
\ben
\Phi=\Phi_0=const,\,\,\, g_{\mu\nu}=\eta_{\mu\nu}
\la{TVSol}
\een
from dynamical equations (\ref{FEq}) we obtain:
\ben
\Phi_0 {\frac{d\Pi}{d\Pi}}(\Phi_0)+\Pi(\Phi_0)=0 \hskip 3.truecm\\
T_{\mu\nu}^0=-{\frac {c^2}{\bar\kappa}}\Lambda U_0 g_{\mu\nu}=
TT_{\mu\nu}^0- <\rho_0>c^2 g_{\mu\nu},
\la{TVEq}
\een
where $U_0=\Phi_0\Pi(\Phi_0)=\Phi_0\Pi_0$. But for true vacuum
solution we must have  \ben TT_{\mu\nu}^0\equiv 0.
\la{TVacuumTT}\een This way we obtain
\ben
 <\rho_0>={\frac 1 {\bar\kappa}}\Lambda U_0={\frac 1 \kappa}\Lambda \Pi_0
\la{rho}
\een

\newpage
Hence in MDG: \vskip 3.truecm
 \centerline{ \hskip 1.truecm True Vacuum \hskip
1.truecm $\Rightarrow $ \hskip 1.truecm Minkowski Space-Time:}
\vskip -1.truecm  $$TT_{\mu\nu}\equiv 0.$$
 \vskip 2.truecm
\centerline{Physical Vacuum \hskip 1.truecm $\Rightarrow $ \hskip
1.truecm  de Sitter Space-Time:}  \vskip -1.truecm $$TT_{\mu\nu}=
<\!\rho_0\!>c^2 g_{\mu\nu}.$$

 \vskip 2.truecm

 {\bf -- a physically sound picture !}

 \vskip 1.truecm
 The real word looks like de Sitter Universe created by zero quantum
vacuum fluctuations and perturbed by other matter and radiation
fields.

\newpage
For $<\rho_0>$ calculated using Plank length as a quantum cuttoff
the observed value of $\Lambda$ gives:

$$U(\Phi_0)/U(\bar\Phi)=\kappa <\rho>/\Lambda \approx 10^{122}$$

and causes the famous {\bf cosmological constant problem} in
standard theory. In different articles this number varies from
$10^{118}$ to $10^{123}$.

We see that:

 $\bullet$ It is obviously close in order to the ratio
of cosmological action ${\cal A}_c$ and Planck constant $\hbar$:
$$U(\Phi_0)/U(\bar\Phi)\approx {\cal A}_c/\hbar.$$

$\bullet$  In MDG there is no crisis caused by this big number,
because it gives ratio of the values of cosmological potential for
different solutions: $\Phi_0$ and $\bar\Phi$, i.e. {\bf in
different universes}.

\newpage
If we calculate the values $${\cal A}_{G,\Lambda}^0= -\Lambda\,
{\frac c
 {\bar\kappa}}\, U_0 \,\, V\!oll$$ and $$\bar{\cal
A}_{G,\Lambda}=\Lambda\,{\frac c {\bar\kappa}}\, \bar U \,\,
V\!oll $$ of the very action (\ref{A_Gc}) and introduce
corresponding specific actions \,$\alpha_0=-\Lambda\, {\frac c
 {\bar\kappa}}\, U_0$\, and
\,$\bar\alpha=\Lambda\,{\frac c {\bar\kappa}}\, \bar U$,\, i.e.,
actions per unit volume, we can rewrite the above observed result
in a form:

$$\bar\alpha \approx  -\alpha_0  \times \hbar /{\cal A}_c=
|\alpha_0| \times 10^{-122}. $$

One can hope that such new and quite radically changed formulation
of the cosmological constant problem in MDG will bring us to its
resolution. For example it's easy to think that this results is
determined by evolution of the Universe.

\newpage
Indeed, consider the simplest model of Universe build of Bohr
hydrogen atoms  in ground state only, i.e. let's just for
simplicity describe the whole content of the Universe with such
{\bf effective Bohr hydrogen} (EBH) atoms.

Then for the  time of the existence of Universe $T_U \sim 4\times
10^{17} sec$ one EBH atom with Bohr angular velocity $\omega_B =
m_e e^4\hbar^{-3}\sim 4\times 10^{16} sec^{-1}$ accumulates
classical action $${\cal A}_{EBH}= 3/2 \,\omega_B T_U\, \hbar \sim
2.4\times 10^{34} \,\hbar.$$ Hence, to explain the present day
action of the Universe $\sim {\cal A}_c$ the number of EBH in it
must be $$N_{EBH}\sim 5\times 10^{87}$$ which seems to be quite
reasonable number from physical point of view, taking into account
that the observed number of barions in the observable Universe is
$$N_{barions}\sim 10^{78}$$.

\newpage
This means that in our approach we have disposable some $9$ orders
of magnitude to solve cosmological constant problem taking into
account the contribution to the action of Universe of all other
constituents of matter and radiation (quarks , leptons, gamma
quanta, etc) during the evolution of Universe from Big Bang to
present day state.

The main conclusion of this qualitative consideration is that
actually in MDG  {\bf the observed nonzero value of cosmological
constant $\Lambda^{obs}\neq 0$ restricts the number of degrees of
freedom in the observable Universe} and forbids existence of much
more levels of matter structure below the quark level.

\newpage

{\bf\large Application of MDG  in Cosmology}

Consider RW adiabatic homogeneous isotropic Universe with
$$ds^2_{RW}= c^2 dt^2 - A^2 dl^2_k,$$ $t=T_c\tau$, $A(t)=A_c
a(\tau)$ and dimensionless $$dl^2_k={\frac
{dl^2}{1-kl^2}+l^2(d\theta^2+sin^2\theta)d\varphi^2}$$ ($k=-1, 0,
1$) in presence of matter with energy-density
$\varepsilon=\varepsilon_c\epsilon(a)/\bar\Phi$ and pressure
$p=\varepsilon_c p_\epsilon(a)/\bar\Phi$.

Basic dynamical equations of MDG for RW Universe are:
\ben
{\sfrac 1 a}{\sfrac {d^2\!a}{d\tau^2}}+
{\sfrac 1 {a^2}}({\sfrac {da}{d\tau}})^{{}_2}+{\sfrac k {a^2}}=
{\sfrac1 3} \left(\Phi{\sfrac{d\Pi}{d\Phi}}(\Phi)+\Pi(\Phi)\right),\nonumber\\
{\sfrac 1 a}{\sfrac {da}{d\tau}}{\sfrac {d\Phi}{d\tau}}+
\Phi \left({\sfrac 1 {a^2}}({\sfrac {da}{d\tau}})^{{}_2}
+{\sfrac k {a^2}}\right)={\sfrac1 3}\left(\Phi\Pi(\Phi)+ \epsilon(a)\right).
\la{DERWU}
\een

The use of Hubble parameter $h(a)=a^{-1}{\sfrac
{da}{d\tau}}(\tau(a))$ ($\tau(a)$ -- inverse function of
$a(\tau)$), new variable $\lambda=\ln a$ and prime for
differentiation with respect to $\lambda$ gives the system for
$\Phi(\lambda)$ and $h^2(\lambda)$:
\ben
{\sfrac 1 2}(h^2)^\prime +2 h^2 + k e^{-2\lambda}= {\sfrac 1
3}\left(\Phi{\sfrac{d\Pi}{d\Phi}}(\Phi)+\Pi(\Phi)\right),\nonumber
\\ h^2 \Phi^\prime +\left(h^2+k e^{-2\lambda}\right)\Phi= {\sfrac
1 3}(\Phi\Pi(\Phi)+\epsilon(e^\lambda)).\nonumber
\la{NDE}
\een
and relation
\ben
\tau(a)=\int^a_{\!a_{in}}\!da /(a\,h(a))+\tau_{in}.\nonumber
\een
Excluding cosmological factor $\Pi(\Phi)$ we have:

\ben
\Phi^{\prime\prime}\!+\!
\left({\sfrac {h^{\prime}} h}\!-\!1\right)\Phi^{\prime}
\!+\!2\left({\sfrac {h^{\prime}} h}\!-
k h^{-2}e^{-2\lambda}\right)\Phi
\!=\!{\sfrac 1 {3h^2}}\epsilon^{\prime}.
\la{ODEPhi}
\een

In terms of the function $\psi(a) = \sqrt{|h(a)|/a}\,\Phi(a)$ it
reads:
\ben
\psi^{\prime\prime} + n^2 \psi = \delta,
\la{DEPsi}
\een
where we introduce new functions
\ben
-n^2 = {\sfrac 1 2}{\sfrac {h^{\prime\prime}} h}-
{\sfrac 1 4}({\sfrac {h^{\prime}} h})^2\!-
{\sfrac 5 2}{\sfrac {h^{\prime}} h}+{\sfrac 1 4}
+{\sfrac {2 k} {h^2}} e^{-2\lambda},\nonumber \\
\delta = {\sfrac 1 3}\sqrt{{ a/{|h|^3}}}
{\sfrac {d\epsilon}{da}} .
\la{n,delta}
\een

\newpage
Now we are ready to consider

{\bf The inverse cosmological problem}: to find a cosmological
factor $\Pi(\Phi)$ (or potentials $V(\Phi)$, or $U(\Phi)$) which
yield given evolution of the Universe, determined by function
$a(\tau)$.

{\em A remarkable property of MDG:}

{\bf An unique solution of this problem exist for almost any three
times differentiable function $a(\tau)$.}

\newpage

Indeed: for given $a(\tau)$ construct a function $h(\lambda)$ and
find the general solution $\Phi(\lambda,C_1,C_2)$ of the {\em
linear} second order differential equation (\ref{ODEPhi}). The two
constants $C_{1,2}$ have to be determined from the additional
conditions $$\Pi(\bar\Phi)=1,\,\,\, {\sfrac
{d\Pi}{d\Phi}}(\bar\Phi)=\bar\Phi^{-1},\,\,\, {\sfrac
{d^2\Pi}{d\Phi^2}}(\bar\Phi)={\sfrac 3 2}p^{-2}\bar\Phi^{-2}.$$
--self-consistence conditions at point $\bar\lambda$ which is real
solution of the algebraic equation $$r(\bar\lambda)=-4,$$
$$r(\lambda)=-6\left({\sfrac 1 2}(h^2)^\prime +2 h^2 +k
e^{-2\lambda}\right)$$ \noindent being dimensionless scalar
curvature: $r= R/\Lambda$. Then: \ben \bar\Phi=-4\bar\epsilon
\left(1\!+\!{\sfrac 4 3}p^2\right)/ \!\left(\bar
j_{00}^\prime(1\!+\!{\sfrac 4 3}p^2) + 4p^2\bar h^2\bar r^\prime
\right), \nonumber \\ \bar \Phi^\prime/\bar \Phi = -{\sfrac 1
3}p^2 \bar r^\prime/\left(1\!+\!{\sfrac 4 3}p^2\right).\hskip
4truecm \la{barPhiPhi'} \een Here
$j_{00}=G_{00}/\Lambda=3\left(h^2+ke^{-2\lambda}\right)$ is
dimensionless $00$-component of Einstein tensor. Hence, the values
of all "bar" quantities (including $\bar \kappa$ in action
(\ref{A_Gc})) may be determined from time evolution $a(\tau)$ of
the Universe via the solution $\bar \lambda = \ln \bar a$ of the
equation (\ref{RV}). In their turn $\bar\Phi$ and
$\bar\Phi^\prime$ determine the values of constants $C_{1,2}$ and
an unique solution $\Phi(\lambda)$ of the equation (\ref{ODEPhi}):
$$\Phi(\lambda)=C_1\Phi_1(\lambda)+C_2\Phi_2(\lambda)+
\Phi_\epsilon(\lambda)$$ \noindent where $\Phi_1(\lambda)$ and
$\Phi_2(\lambda)$ are a fundamental system of solutions of the
homogeneous equation associated with non-homogeneous one
(\ref{ODEPhi}). Then $$\Phi_\epsilon\!=\!{\frac{\bar a}{(3\bar
h\bar\Delta)}}\!\left(
\Phi_2\!\int_{\bar\lambda}^\lambda\!d\epsilon\,{\frac
{\Phi_1}{ah}}- \Phi_1\!\int_{\bar\lambda}^\lambda\!d\epsilon\,
{\frac {\Phi_2} {ah} } \right).$$ where $\Delta(\lambda)=
\Phi_1\Phi_2^\prime-\Phi_2\Phi_1^\prime$. The cosmological factor
$\Pi$ and the potential $V$ as functions of the variable $\lambda$
are determined by equations \ben \Pi(\lambda) = j_{00} + 3 h^2
\Phi^\prime/\Phi - \epsilon/\Phi, \nonumber \\ V(\lambda)={\sfrac
2 3} \int\Phi\left(\Phi\Pi^\prime-\Phi^\prime\Pi \right)d\lambda
\la{PiV} \een which define functions $\Pi(\Phi)$ and $V(\Phi)$
implicitly, too.

\newpage
{\bf Simple exactly soluble examples:}

1. {\bf Evolution law} $a(\tau)\!=\!(\omega\tau)^{1/\nu}$,
($\omega$ is free parameter) gives $$h(\lambda)\!=\!{\sfrac \omega
\nu}e^{-\nu\lambda},\,\,\, -n^2(\lambda)=
{\sfrac{1+10\nu+\nu^2}4}+2k{\sfrac{\nu^2}{\omega^2}}e^{2(\nu-1)\lambda},$$
 and the equation
$\bar a^2\!+{\sfrac {3(\nu-2)\omega^2}{2\nu^2}}\bar a^{2(1-\nu)}\!=\!k$
for $\bar a$.

i) For $\nu\geq 2$ we have real solution $\bar a$ only if $k=+1$:
$$
\Phi_1(a)=a^{{\frac {\nu+1}2}}I_\mu\!\left(b a^{\nu-1}\right),
\Phi_2(a)=a^{{\frac {\nu+1}2}}K_\mu\!\left(b a^{\nu-1}\right),
$$
$\mu:={ \sfrac {\sqrt{1+10\nu+\nu^2}} {2(\nu-1)} } $ being the order
of Bessel functions $I_\mu,J_\mu,K_\mu,Y_\mu$ and
$b:=\sqrt{{\sfrac {2\nu^2}{(\nu-1)^2\omega^2}}}$.
For GR-like law $a\!\sim\!\sqrt{\tau}$ ($\nu\!=\!2$) in MDG
we obtain positive value ($k=+1$) for three-space curvature,
$\bar\lambda\!=\!0$, $\mu={\sfrac 5 2}$
and Bessel functions are reduced to elementary functions.

ii) When $\nu<2$ all values $k=-1,0,+1$ are admissible:

- for $k\!=\!+1$ the solutions $\Phi_{1,2}$ are the same as above;

- for $k\!=\!0$ we have $$\Phi_{1,2}=a^{{\frac
{\nu+1}2}\pm\mu(\nu-1)};$$

- for $k\!=\!-\!1$ the solutions are: $$ \Phi_1\!=\!a^{{\frac
{\nu+1}2}}J_\mu\!\left(ba^{\nu-1}\right), \Phi_2\!=\!a^{{\frac
{\nu+1}2}}Y_\mu\!\left(ba^{\nu-1}\right). $$ In the special case
of linear evolution $a\!\sim\!\tau$ ($\nu=1$)
$-n^2(\lambda)=3+{\sfrac {2k}{\omega^2}}$,\, $$\Phi_{1,2}(a)=
a^{1\pm \sqrt{-n^2}}$$ \noindent and the root
$\bar\lambda\!=\!{\sfrac 1 2}\ln{\sfrac 3 2}(k+\omega^2)$ is real
for all values of $\omega^2\!>\!0$, if $k\!=\!0,+1$. For
$k\!=\!-1$ the root $\bar\lambda$ will be real if
$|\omega|\!>\!1$.

\newpage
2. {\bf Evolution law} $a(\tau)\!=\!e^{\omega\tau}$ gives
$$h(\lambda)\!=\!\omega,\,\,\,\, -n^2(\lambda)\!=\!{\sfrac 1
4}+{\sfrac {2k}{\omega^2}}e^{-2\lambda}$$ and the equation
${\sfrac 2 3}\bar a^2 (1-3\omega^2)\!=\!k$ with root $\bar
a\!=\!\sqrt{{\sfrac 3 {2|1-3\omega^2|}}}$. Now we have the
following solutions:

i) $|\omega|\!<\!\!{\sfrac{\sqrt{3}}3}$:
$\Phi_1\!=\!a\cosh\!\left(\!{\sfrac{\sqrt{2}}{|\omega|a}}\!\right)$,
$\Phi_2\!=\!a\sinh\!\left(\!{\sfrac{\sqrt{2}}{|\omega|a}}\!\right)$,
$k\!=\!+1$;

ii)$|\omega|\!>\!\!{\sfrac{\sqrt{3}}3}$:
$\Phi_1\!=\!a\,\cos\!\left(\!{\sfrac{\sqrt{2}}{|\omega|a}}\!\right)$\,\,,\,
$\Phi_2\!=\!a\,\sin\!\left(\!{\sfrac{\sqrt{2}}{|\omega|a}}\!\right)$\,,
$k\!=\!-1$.

Conditions  $r(\bar\lambda)\!=\!-4$ and (\ref{barPhiPhi'}) exclude
{\em exact} exponential expansion of spatially flat Universe ($k\!=\!0$ ).
For inflationary scenario in this case one may use a scale factor
$a(\tau)\!=\!e^{\omega\tau}\!+\!const$ which turns to be possible if $const\!\neq\!0$.

\newpage
{\bf Specific properties of MDG:}

1) If $n\!>\!0$ {\bf dilatonic field $\Phi(a)$ oscillates}; if
$n\!<\!0$ such oscillations do not exist. Dilatonic field $\Phi$
may change its sign, i.e. {\bf phase transitions of the Universe
from gravity ($\Phi\!>\!0$) to anty-gravity ($\Phi\!<\!0$) and
vice-versa} are possible in general for width class of
cosmological potentials.

2) In spirit of Max principle {\bf Newton constant depends on
presence of matter:} $G\!\sim\!1/\bar\Phi\!\sim\!1/\bar\epsilon$.

3) For simple functions $a(\tau)$ the cosmological factor
$\Pi(\Phi)$ and potentials $V(\Phi)$ and $U(\Phi)$ may show
unexpected {\bf catastrophic behavior:}
$$\sim\!(\Delta\Phi)^{3/2}$$
($\Delta\Phi\!=\!\Phi\!-\!\Phi(\lambda^\star)$) in vicinity of the
critical points $\lambda^\star$:
$\Phi^\prime(\lambda^\star)\!=\!0$ of the projection of analytical
curve $\{\Pi(\lambda),\Phi(\lambda),\lambda\}$ on the plain
$\{\Pi,\Phi\}$.

Scale factors $a(\tau)$ exist yielding an {\bf everywhere
analytical cosmological factor $\Pi(\Phi)$ and potentials
$V(\Phi)$ and $U(\Phi)$}, too.

4) Clearly one can construct {\bf MDG model of Universe without
initial singularities:} $a(\tau_0)=0$ (typical for GR) and with
{\bf any desired kind of inflation.}

5) Because the dilaton field $\Phi$ is quite massive, in it will
be stored significant amount of energy. An interesting open
question is: {\bf may the field $\Phi$ play the role of dark
matter} in the Universe?

 A very important problem is {\bf to
reconstruct the cosmological factor $\Pi(\Phi)$ of {\em real}
Universe} using proper experimental data and astrophysical
observations.

Maybe the best way to study SUSY breaking and the corresponding
potential $V_{SUSY}(\phi)$ is to look at the sky and to try to
reconstruct the {\bf real time evolution} of the Universe.

\newpage
{\bf GENERAL CONCLUSIONS:}

$\bullet$ MDG is a rich model which offers new curious
possibilities and deserves further careful investigation.

$\bullet$ {\bf Instead of  Cosmological Constant we have to prefer
a  Cosmological Potential}, because it yields much richer theory
and may be more suitable for description of the real world.

\newpage

{\bf Acknowledgments:}

The author is deeply indebted to NORDITA and personally to
Professor  Holger Bech Nielsen, Dr. Kimmo Kainulainen and Ellen
Pedersen for the help and support which make possible his
participation in the conference.

\end{document}